\documentstyle[aps]{revtex}
\begin{document}
\draft
\title{Kinetic model of three component, weakly ionized,
collisional plasma with a beam of neutral particles}
\author{David Tsiklauri}
\address{Physics Department, Tbilisi State University,
3 Chavchavadze Ave., Tbilisi 380028, Georgia;
email: dtsiklau@usa.net}
\maketitle
\begin{abstract}
Kinetic model of three component, weakly ionized, collisional
plasma with a beam of neutral particles is developed.
New dispersion relations for linear perturbations are derived
and analyzed in various limiting cases.
\end{abstract}

\section{Introduction}

It is well-known that
neutral beam injection is one of the fundamental fusion plasma
heating methods.  In general, a particle
accelerator is used to create
fast ion beams (the particle energies are on the order of
100 keV); the ion beam is then passed through a neutral gas
region, where the ions neutralize via charge-exchange reactions
with the neutral gas. The neutralized beam is then injected into
a magnetically confined plasma.  Of course, the neutral atoms
are unaffected (not confined) by the magnetic field, but ionize
as they penetrate into the plasma.  Then the high-energy ions
transfer fraction of their energy to the plasma particles in
repeated collisions, and heat the plasma \cite{1,2,3,4,5,6}.

In this paper we develop a kinetic model of three component,
weakly ionized, collisional plasma with a beam of neutral
particles. We employ a kinetic equation for the charged
particles of $\alpha$ sort in the weakly ionized plasma
with the Batnagar-Gross-Krook (BGK) model collisional term.
Similar model has been developed previously by others \cite{abr}.
In this book authors do not take into account possibility of
existence of regular velocity of the neutral particles
\cite{abr}.
In the light of the possible relevance of our model for the
heating of plasma by neutral beam injection, we set out with
the aim to generalize results of Ref. \cite{abr} by
allowing neutral particles to have regular velocity
and seek for possible novelties brought about by
this effect. Indeed, the dispersion relations for linear
perturbations obtained in this paper differ substantially from
those of Ref. \cite{abr}.

In section II we formulate our model and obtain general
dispersion relation. In section III we analyze various limiting
cases of the dispersion relation and discuss the results.

\section{The model}

We start analysis of the dielectric permittivity (DP) of
a collosional plasma with weakly ionzed, non-degenerate
plasma when the integral of elastic collisions in the
kinetic equation for the charged particles
can be apporximated by the BGK term, while
it is possible to neglect the collisions between the
chaged particles themselvs.
Analysis of this relatively simple model will be useful
for further more complicated case of fully ionized
plasma (which, in fact, is more relevant for the fusion
plasma). The latter is beyond the scope of present paper
and the separate analysis needs to be done.

The kinetic equation for the charged particles
of $\alpha$ sort in the weakly ionized plasma with the
BGK model collisional term
can be written as following \cite{abr}:

$$
{{\partial f_\alpha}\over{\partial t}}+
\vec v \cdot {{\partial f_\alpha}\over{\partial \vec r}}
+e_\alpha\{ \vec E + \vec v \times \vec B \}
{{\partial f_\alpha}\over{\partial \vec p}}
= - \nu_{\alpha n} (f_\alpha -N_\alpha
\Phi_{\alpha n}).
\eqno(1)
$$
Here, $\nu_{\alpha n}$ denotes collision frequency of
charged particles with the neutrals, which in this model
is assumed being constant, whereas
$$
N_\alpha \equiv \int d \vec p f_\alpha,
$$
and
$$
\Phi_{\alpha n}\equiv{{1}
\over{(2 \pi m_\alpha T_{\alpha n})^{3/2}}}
\exp{\left[-m_\alpha (\vec v - \vec V_0)^2/
(2 T_{\alpha n})\right]},
\;\;\; T_{\alpha n} \equiv {{m_\alpha T_n +
M_n T_\alpha}\over{m_\alpha + M_n}}.
\eqno(2)
$$
Index $\alpha$ ($\alpha= e,i$) refers to charged particles
(electrons and ions), whereas $n$ --- to neutrals.
$\vec V_0$ denotes regular, uniform velocity of the neutral particles.
Finally, $T_\alpha$ is defined by following expression:
$$
T_\alpha = {{m_\alpha}\over{2 N_\alpha}} \int d \vec p
(\vec v - \vec V_\alpha)^2 f_\alpha
$$

The specific form of the BGK integral used here is derived from
its more general form \cite{abr}
$$
{\left({{\partial f_\alpha}\over{\partial t}}
\right)}^{\alpha \beta}_{BGK}=
- \nu_{\alpha \beta}(f_\alpha - N_\alpha \Phi_{\alpha \beta}),
\eqno(3)
$$
where $\nu_{\alpha \beta}$ is some constant which has meaning
of effective collision frequency between particles of $\alpha$
and $\beta$ sort, i.e. it characterizes time of momentum
relaxation of $\alpha$ sort particles as a result of their
collision with particles of $\beta$ sort. Function
$\Phi_{\alpha \beta}$ is determined by following expression:
$$
\Phi_{\alpha \beta}\equiv{{1}\over{(2 \pi m_\alpha
T_{\alpha \beta})^{3/2}}}
\exp{\left[-m_\alpha (\vec v - \vec V_\beta)^2/(2
T_{\alpha \beta})\right]},
\eqno(4)
$$
here $V_\beta=(1/N_\beta) \int d \vec p \, \vec v f_\beta$.

It should be emphasized that the BGK collisonal integral
describes accurately collisions only particles of different
sort. Therefore, it can be used to describe collisions of
charged paricles with the neutrals in weakly ionized plasma,
when the scatteing of charged particles on the neutrals is a
dominant process. In the case of fully ionized plasma, in spite
of its relative simplicity, use of BGK integral is not
justified \cite{abr}.

In what follows, we consider isothermal models of the
BGK integral, i.e. we neglect change in temperature of
charged particles with chage in their corresponding distribution
functions. We ought to mention that the results obtained here
will be qualitatively the same for the non-isothermal model of
BGK integral. We further assume that the masses and the
temperatures of the ions and neutrals do coincide, i.e.
$m_i=M_n \equiv M$ and $T_i=T_n$. In this case to the order of
$\sim m_e/M$ terms we have $T_{e n}=T_e$.
Thus, in the Eq.(2), under these simplifying assumptions we can
set $T_{\alpha n}=T_\alpha$ and
$$
\Phi_{\alpha n}={{1}\over{(2 \pi m_\alpha T_\alpha)^{3/2}}}
\exp{\left[-m_\alpha (\vec v- \vec V_0)^2/(2 T_\alpha)\right]}
\eqno(5)
$$
which, in fact, coincides with the Maxwellian distribution
function (with the beam having velocity $\vec V_0$) normalized
to unity.

In the static equilibrium state, with the external fields absent,
Eq.(1) allows for  the only solution
$f_{0 \alpha}=N_{0 \alpha} \Phi_{0 n}$. In what follows
subscript 0 will denote unperturbed and $\delta$ perturbation
of the physical quantities.

Let us consider small perturbation of the distribution function
$\delta f_\alpha$ which is caused by appearance of small fields
$\vec E$ and $\vec B$. After usual linearization of the Eq.(1)
we obtain
$$
{{\partial \delta f_\alpha}\over{\partial t}}+
\vec v \cdot {{\partial \delta f_\alpha}\over{\partial \vec r}}
+e_\alpha \vec E \cdot
{{\partial f_{0 \alpha}}\over{\partial \vec p}}
= - \nu_{\alpha n} (\delta f_\alpha - \int d \vec p
\delta f_\alpha \Phi_{\alpha n}).
\eqno(6)
$$

The solution of the latter equation for the plane monochromatic
waves (i.e. $\vec E, \delta f_\alpha \sim \exp{\left[
-i \omega t + i \vec k \cdot \vec r\right]}$) can be written as
$$
\delta f_\alpha = i {{e_\alpha}\over{T_\alpha}}
{{f_{0 \alpha} \left[{\vec v \cdot \vec E - \vec V_0 \cdot
\vec E}\right]}\over{\omega + i \nu_{\alpha n} -
\vec k \cdot \vec v}}
+ {{i \nu_{\alpha n} \eta_\alpha f_{0 \alpha}}\over{\omega +
i \nu_{\alpha n} - \vec k \cdot \vec v}},
\eqno(7)
$$
where $\eta_\alpha=(1/N_{0 \alpha}) \int d \vec p
\delta f_\alpha$,
which is perturbation of the particle number density normalized
to equilibrium value of the number dinsity.
$\eta_\alpha$ can be calculated either by integration of the
Eq.(7) over momentum or by using the continuity equation
for the particles of $\alpha$ sort:
$$
\eta_\alpha ={{\vec k \cdot \vec j_\alpha}\over{e_\alpha
N_{0 \alpha} \omega}},
$$
$$
\vec j_\alpha = e_\alpha
\int d \vec p \, \vec v \delta f_\alpha ,
\eqno(8)
$$
here, $\vec j_\alpha$ denotes charge current of
particles of $\alpha$ sort.

It is known that the complex tensor of DP can be written as
$$
\varepsilon_{ij}(\omega, \vec k)= \delta_{ij}+
{{i}\over{\varepsilon_0 \omega}} \sigma_{ij}(\omega, \vec k)
\eqno(9)
$$
where $\delta_{ij}$ is usual Kroneker tensor and
$\sigma_{ij}(\omega, \vec k)$ is the conductivity tensor
defined by
$$
j_{i}= \sum_\alpha j_{i \alpha}=\sigma_{ij}
(\omega, \vec k) E_j.
\eqno(10)
$$

In general when $\varepsilon_{ij}$ tensor is of the
type $\varepsilon_{ij}= \delta_{ij}+A_iA_j-A_iB_j$,
then defining quantities $\varepsilon^l$ and
$\varepsilon^{tr}$ as
$$
\varepsilon^l= {{k^ik^j}\over{k^2}} \varepsilon_{ij}=
1+{{(\vec k \cdot \vec A)^2}\over{k^2}}-
{{(\vec k \cdot \vec A)
(\vec k \cdot \vec B)}\over{k^2}}
\eqno(11)
$$
and
$$
\varepsilon^{tr}= {{1}\over{2}}
\left(\delta^{ij} - {{k^ik^j}\over{k^2}} \right)
\varepsilon_{ij}=
1+{{(\vec k \times \vec A)^2}\over{2 k^2}}-
{{(\vec k \times \vec A)
(\vec k \times \vec B)}\over{2 k^2}}
\eqno(12)
$$
respectively, we can split tensor from Eq.(9) in the
longitudinal and transverse (with respect to wave-vector
$\vec k$) parts as following:
$$
\varepsilon_{ij}(\omega, \vec k)=
\left(\delta_{ij} - {{k_ik_j}\over{k^2}} \right)
\varepsilon^{tr}(\omega, k) +
{{k_ik_j}\over{k^2}} \varepsilon^l(\omega, k) \eqno(13)
$$

Now, inserting expression for $\delta f_\alpha$ from the
Eq.(7) into Eq.(8) and using Eqs.(10)-(12)
we obtain following expessions for
$\varepsilon^{l}$ and $\varepsilon^{tr}$:
$$
\varepsilon^l= 1 - \sum_\alpha {{\omega^2_{L \alpha}}
\over{k^2 \omega}}
{{1}\over{(2 \pi)^{3/2}}}{{1}\over{V^5_{T \alpha}}}
\left[{
\int d \vec v {{(\vec k \cdot \vec v)^2
e^{-v^2/(2V^2_{T\alpha})}}\over{\omega + i \nu_{\alpha n}-
\vec k \cdot \vec v}}
-
\int d \vec v {{(\vec k \cdot \vec v)(\vec k \cdot \vec V_0)
e^{-v^2/(2V^2_{T\alpha})}}\over{\omega + i \nu_{\alpha n}-
\vec k \cdot \vec v}}
}\right]
\times
$$
$$
\left[
1-{{i \nu_{\alpha n} k_i}\over{\omega}}
{{1}\over{(2 \pi)^{3/2}V^3_{T \alpha}}}
\int d \vec v {{v_i e^{-v^2/(2V^2_{T\alpha})}}
\over{\omega + i \nu_{\alpha n}- \vec k \cdot \vec v}}
\right]^{-1}, \eqno(14)
$$
$$
\varepsilon^{tr}= 1 - \sum_\alpha {{\omega^2_{L \alpha}}
\over{2 k^2 \omega}}
{{1}\over{(2 \pi)^{3/2}}}{{1}\over{V^5_{T \alpha}}}
\left[{
\int d \vec v {{(\vec k \times \vec v)^2
e^{-v^2/(2V^2_{T\alpha})}}\over{\omega + i \nu_{\alpha n}-
\vec k \cdot \vec v}}
-
\int d \vec v {{(\vec k \times \vec v)(\vec k \times \vec V_0)
e^{-v^2/(2V^2_{T\alpha})}}\over{\omega + i \nu_{\alpha n}-
\vec k \cdot \vec v}}
}\right].  \eqno(15)
$$
Here, $\omega_{L \alpha}=
\sqrt{(e^2N_\alpha)/(\varepsilon_0m_\alpha)}$ and
$V_{T \alpha}=\sqrt{T_\alpha/m_\alpha}$.
The integrals in the Eqs.(14) and (15) may be evaluated by
choosing the $z$-axis along $\vec k$. The integration over
$v_x$ and $v_y$ is elementary. Whereas, $v_z$ integral may be
expressed in terms of a single transcendental function,
which called the plasma dispersion function.
There are several different definitions of this function
used in the literature. We use the one given by
Melrose \cite{mel}:
$$
\bar \phi(z) = - {{z}\over{\sqrt{\pi}}}
\int^{+ \infty}_{- \infty} {{dt e^{-t^2}}\over{t-z}}.
\eqno(16)
$$

Using Eq.(16) and following intermediate results of
integration
$$
\int d \vec v {{(\vec k \cdot \vec v)^2
e^{-v^2/(2V^2_{T\alpha})}}\over{\omega + i \nu_{\alpha n}-
\vec k \cdot \vec v}}= -
\sqrt{\pi} z [\bar \phi(z)-1], \eqno(17)
$$
$$
\int d \vec v {{(\vec k \cdot \vec v)(\vec k \cdot \vec V_0)
e^{-v^2/(2V^2_{T\alpha})}}\over{\omega + i \nu_{\alpha n}-
\vec k \cdot \vec v}}
= (2 \pi)^{3/2} V^3_{T \alpha}(\vec k \cdot \vec V_0)
[\bar \phi(z) -1], \eqno(18)
$$
$$
{{i \nu_{\alpha n} k_i}\over{\omega}}
{{1}\over{(2 \pi)^{3/2}V^3_{T \alpha}}}
\int d \vec v {{v_i e^{-v^2/(2V^2_{T\alpha})}}
\over{\omega + i \nu_{\alpha n}- \vec k \cdot \vec v}}
={{i \nu_{\alpha n}}\over{\omega}}[\bar \phi(z) -1], \eqno(19)
$$
$$
\int d \vec v {{(\vec k \times \vec v)^2
e^{-v^2/(2V^2_{T\alpha})}}\over{\omega + i \nu_{\alpha n}-
\vec k \cdot \vec v}}=
{{\sqrt{\pi}}\over{zk}}\bar \phi(z), \eqno(20)
$$
$$
\int d \vec v {{(\vec k \times \vec v)(\vec k \times \vec V_0)
e^{-v^2/(2V^2_{T\alpha})}}\over{\omega + i \nu_{\alpha n}-
\vec k \cdot \vec v}} =0, \eqno (21)
$$
where, $z= (\omega + i \nu_{\alpha n})/(\sqrt{2}k V_{T \alpha})$,
we obtain
$$
\varepsilon^l=1+ \sum_\alpha {{\omega^2_{L \alpha}}
\over{k^2 V^2_{T \alpha}}}
{{[1- \bar \phi(z)][1-(\vec k \cdot \vec V_0)/
(\omega +i \nu_{\alpha n})]}\over{1-[(i \nu_{\alpha n})/
(\omega + i \nu_{\alpha n})] \bar \phi(z)}}, \eqno(22)
$$
$$
\varepsilon^{tr}=1- \sum_\alpha {{\omega^2_{L \alpha}}
\over{\omega(\omega + i \nu_{\alpha n})}} \bar \phi(z).
\eqno(23)
$$
Note, that conventinal kinetic model of three component, weakly
ionized, collisional plasma \cite{abr} is significantly
modified by taking into account possible existence of
a beam of neutral particles. Namely, the expression for the
$\varepsilon^l$ is modified by additional factor
$[1-(\vec k \cdot \vec V_0)/(\omega +i \nu_{\alpha n})]$.
While the form of the $\varepsilon^{tr}$ is not changed by
the presence of the beem.

\section{Discussion}

Let us start analysis of the obtained results from
longitudinal waves as we have seen that transverse waves
do not incur any modification by the presence of the beem
of neutral particles. The dispersion relation for the
longitudinal waves reads as following:
$$
\varepsilon^l=1+ \sum_\alpha {{\omega^2_{L \alpha}}
\over{k^2 V^2_{T \alpha}}}
{{[1- \bar \phi(z)][1-(\vec k \cdot \vec V_0)/
(\omega +i \nu_{\alpha n})]}\over{1-[(i \nu_{\alpha n})/
(\omega + i \nu_{\alpha n})] \bar \phi(z)}}=0 \eqno(24)
$$
The latter equation is a transcendental one, thus, in general
case, it has many complex solutions $\omega(k)$. Let us
consider the most interesting ones which correspond to weakly
damped oscillations.

Let us consider, first, high frequency waves, i.e. when
$\omega \gg k v_{T \alpha}, \nu_{\alpha n}$.
Using asymptotic expansion for $\bar \phi(z)$ \cite{mel}
$$
\bar \phi(z)=1+{{1}\over{2z^2}}+{{3}\over{4z^4}}+ ...-
i \sqrt{\pi}ze^{-z^2}, \;\;\; {\rm when} \;\;\;
|z| \gg 1\eqno(25)
$$
we obtain following dispersion relation for the weakly
damped waves (Re $\omega \gg$ Im $\omega$)
$$
\varepsilon^l= 1 -\left[{{\omega^2_{Le}}\over{\omega^2}}
\left({1+{{3 k^2 V^2_{Te}}\over{\omega^2}}}\right)
-i\left\{ \sqrt{{\pi}\over{2}}{{\omega \omega^2_{Le}}
\over{k^3V^3_{Te}}} \exp{\left[{-{{\omega^2}
\over{2k^2V^2_{Te}}}}\right]} + {{\omega^2_{Le} \nu_{en}}
\over{\omega^3}}\right\} \right]
\left[{1-{{\vec k \cdot \vec V_0}\over{\omega}}}\right]=0.
\eqno(26)
$$
Here, we neglect the contribution from ions, because
it is significant when $T_i \geq T_e (M/m_e)^2$, i.e. when
the temperature of ions is greater than the temperature of
electrons by more than six orders of magnitude. It is unlikely
that such differences in the temperatures actually do
realize in the nature \cite{abr}. Therefore, in the
frequency domain concerned, the plasma can be considered
as a purely electronic, i.e. the role of the ions is reduced
only to neutralize the charge of electrons.
The dispersion relation (26) has to imaginary terms.
The first one describes collisionless Cherenkov absorption
of the plasma waves. Whereas, the second one has purely
collisional nature and describes dissipation of the fields
energy in via collisions (electronic friction) \cite{abr}.
The difference induced by the presence of the beam of neutral
particles is presented by a factor
(see, Ref. \cite{abr} for comparison)
$$
\left[{1-{{\vec k \cdot \vec V_0}\over{\omega}}}\right].
\eqno(27)
$$

In addition to the high frequency longitudinal oscillations
in isotropic collsionless plasma there also exist low frequency
oscillations, so called, Ion-acoustic waves. They exist
in highly non-isotermal plasma, where $T_e \gg T_i$.
Phase velocity of these waves lies in the
$V_{Ti} \ll \omega / k \ll V_{Te}$ domain.
It is obvious, that such
waves should also exist in collisional plasma if the
collisions are sufficiently rare. Thus, when
$\omega \ll \nu_{in}$ and $|\omega+i \nu_{en}| \ll k V_{Te}$
in the $V_{Ti} \ll \omega / k \ll V_{Te}$ phase velocity
domain we obtain following dispersion relation
$$
\varepsilon^l= 1 +
\biggl[{{\omega^2_{Le}}\over{k^2V^2_{Te}}}
\left({1+i\sqrt{{\pi}\over{2}}{{\omega}\over{kV_{Te}}} }\right)
-
$$
$$
{{\omega^2_{Li}}\over{\omega^2}}
\left({1+{{3 k^2 V^2_{Ti}}\over{\omega^2}}}\right)
+i\left\{ \sqrt{{\pi}\over{2}}{{\omega \omega^2_{Li}}
\over{k^3V^3_{Ti}}} \exp{\left[{-{{\omega^2}
\over{2k^2V^2_{Ti}}}}\right]} + {{\omega^2_{Li} \nu_{in}}
\over{\omega^3}}\right\} \biggr]
\left[{1-{{\vec k \cdot \vec V_0}\over{\omega}}}\right]=0.
\eqno(28)
$$
In the latter equation we have used also following asymptotic
expansion
$$
\bar \phi(z)=2z^2-{{4}\over{3}}z^4+...-i \sqrt{\pi}ze^{-z^2},
\;\;\; {\rm when} \;\;\; |z| \ll 1
\eqno(29)
$$
to the first order.

Again, we note that the difference induced by the presence
of the beam of neutral particles is presented by a factor given
by factor Eq.(27) (see, Ref. \cite{abr} for comparison).

This concludes presentaion of the kinetic model of
three component,
weakly ionized, collisional plasma with a beam of
neutral particles.
We have generalized the results of Ref. \cite{abr} by
allowing neutral particles to have regular velocity
(i.e. by allowing for the existence of a beam of neutrals).
We have shown that the novel, generalized dispersion relations
for linear perturbations obtained in this paper differ
substantially from those of Ref. \cite{abr}.
Finally, we would like to conclude outlining,
once again, the possible
relevance of our model for the better understanding
of the plasma heating process by a neutral beam injection.

\end{document}